\documentstyle[epsfig, a4]{article}

\begin{document}

 \title{On the interaction between velocity 
 increment and energy dissipation in the turbulent cascade}

\author{Ch.  Renner$^1$, J. Peinke$^1$, R. Friedrich$^2$
\\ $^1$Fachbereich Physik, Universit\"at Oldenburg
\\ $^2$Institut f\"ur theoretische Physik, Universit\"at M\"unster
}

\date{\today} 

\maketitle
\begin{abstract}
    We adress the problem of interactions between the longitudinal
    velocity increment and the energy dissipation rate in fully
    developed turbulence.  The coupling between these two quantities
    is experimentally investigated by the theory of stochastic
    Markovian processes.  The so--called Markov analysis allows for a
    precise characterization of the joint statistical properties of
    velocity increment and energy dissipation.  In particular, it is
    possible to determine the differential equation that governs the
    evolution along scales of the joint probability density of these
    two quantities.  The properties of this equation provide
    interesting new insights into the coupling between energy
    dissipation and velocity incrementas leading to small scale
    intermittency.
    
\end{abstract}

\section{Introduction}

Important details of the complex statistical behaviour of fully
developed turbulent flows are still unknown, cf. \cite{turb}. Especially
the effect of small scale intermittency, i.e. the phenomenon of
finding unexpected frequent occurences of large fluctuations of the
local velocity on small length scales, is an open problem.

It is commonly accepted that small scale intermittency is due to some
kind of cascading process: Kinetic energy which is fed into the flow
by external forces on some (large) scale $L$ is assumed to be in an
equilibrium with the energy being dissipated by viscosity on the
smallest scale $\eta$.  Inbetween the integral length scale $L$ and the
dissipation scale $\eta$, energy is continuously transported towards
smaller scales by the decay of eddies \cite{K41, K62, Richardson}.

The statistical properties of the turbulent cascade are usually
characterized by means of the difference between the velocities at two
points in space separated by the distance $r$, the so--called
longitudinal velocity increment $u(r)$:
\begin{equation}
	u(r) = {\mathbf e} \left[ {\mathbf v}\left( {\mathbf x} + r
	{\mathbf e}, t \right) - {\mathbf v}\left( {\mathbf x}, t
	\right) \right] .  \label{InkDef}
\end{equation}
The statistics of $u(r)$ is commonly investigated by means of its
moments $S_{u}^{n}(r) = \left< u(r)^n \right>$, the so-called velocity
structure functions.  In the framework of the cascade picture it is
natural to assume that, for scales $r$ within the inertial range $\eta
\ll r \ll L$, the $S_{u}^{n}(r)$ are functions of the scale $r$ and
the rate of energy transfer $\epsilon$ only: $S_{u}^{n}(r) =
f(\epsilon,r)$.  Assuming a constant rate of energy transfer,
Kolmogorov derived his famous result (furtheron referred to as K41)
for the structure functions: $S_{u}^{n}(r) \propto r^{n/3}$
\cite{K41}.

However, as pointed out by Landau \cite{LandauRemark}, there is no
reason to assume that a decaying eddy spreads its energy into equal
parts.  On the contrary, it is likely that the decay of eddies is a
stochastic process.  As a consequence, the energy dissipation rate
$\epsilon(\vec{x},t)$ is a spatially distributed random variable. 
Accordingly, experimental studies yield significant deviations from
Kolmogorov's prediction \cite{turb}.

The statistics of the energy dissipation rate is usually investigated
by the scale--dependence of $\epsilon_{r}$, the average of 
$\epsilon$ over a ball with radius $r$ located at $\vec{x}$:
\begin{equation}
	\epsilon_{r}(\vec{x}) = \frac{1}{V_{\mathcal B} }
	\int\limits_{{\mathcal B}(\vec{x},r) } \epsilon(\vec{x}')
	\; d^{3}x' .
	\label{EpsRDef}
\end{equation}
Taking into account the stochastic nature of $\epsilon_{r}$,
Kolmogorov derived a modified result for the structure functions which
is in better agreement with experimental data (\cite{K62}, furtheron
referred to as K62).

The statstics of the velocity increment can also be characterized by
means of its probability density functions (pdfs) $p(u(r))$.  While
for large scales $r$ the pdfs are almost Gaussian, the phenomenon of
small scale intermittency shows up in a stretched exponential--like
shape of the pdfs on small scales expressing very high probabilities
for large values of $|u(r)|$.  Those deviations from the Gaussian
shape are closely related to the deviations of the structure functions
from the K41 prediction $S_{u}^{n}(r) \propto r^{n/3}$ and can be
attributed to the stochastic nature of the average energy dissipation
rate, see for example $\epsilon_{r}$
\cite{Castaing,Gagne,AntoineBedingt}.

These results clearly show that $\epsilon_{r}$ has an important
influence on the statistics of the velocity increment. It would
therefore be highly desirable to have an experimental tool which
allows for a precise characterization of the interdependence of those
two quantities.  The aim of the present paper is to show that such a
tool is given by the framework of stochastic Markovian processes.

In a recent series of papers \cite{PRLTurb, PhysicaD, JFM,
AntoineEnergie, AntoineLangevin, Greiner}, the Markov analysis has
been applied separately to the velocity increment and the energy
dissipation rate.  The idea, inspired by the cascade picture, is to
consider the velocity increment $u(r)$ (or the energy dissipation
rate $\epsilon_{r}$) as a stochastic variable which evolves in the scale $r$.  If
$u(r)$ fulfills the mathematical condition for a Markov process, the
evolution of $u$ in $r$ can be described by means of the
Fokker--Planck equation, a generalized diffusion equation for the pdf
$p(u(r))$ in the variables $r$ and $u$.  This equation is completely
determined by two coefficients, drift and diffusion coefficient,
respectively, which can be estimated from experimental data.  The
Markov analysis thus provides a possibility to {\it measure} the
stochastic differential equations governing the evolution of the
stochastic variable $u$ in the scale $r$ without incorporating any
models or assumptions on the physics of the systems.  A detailed
explanation of this method is presented in \cite{JFM}.

The mathematics of Markov processes can be generalized to
multidimensional stochastic variables.  In the present paper, we use
the mathematical theory of multidimensional Markovian processes to
establish an unified description of the joint statistical properties
of the velocity increment and the averaged energy dissipation rate. 
Analysing experimental data, we derive a Fokker--Planck equation
describing the evolution of the joint pdf $p(u(r),\epsilon_{r})$ in
the scale $r$.  The properties of this equation provide interesting new
insights into the interdependence of the velocity increment and the
averaged energy dissipation rate.  In particular we differentiate
between deterministic and stochastic coupling.

The paper is organized as follows: First, we shortly summarize the
mathematical formulation of multidimensional Markovian processes in
section \ref{theory}.  Section \ref{eindim} briefly recalls the
results of the one dimensional analysis for the velocity increment and
the averaged energy dissipation rate, respectively.  A short
description of the experimental set--up and the data is presented in
section \ref{setup} while the results of the two dimensional Markov
analysis are given in section \ref{results}.  A short summary and
discussion of our results in section \ref{discussion} will conclude
the paper.

\section{The mathematics of Markov processes}\label{theory}

This section gives a brief summary of the theory of multidimensional
Markovian processes.  For a detailled discussion of the theorems
summarized here, we refer the reader to standard textbooks like
\cite{Risken}.

We consider the two dimensional stochastic variable ${\mathbf q}(r)$
which is defined as:
\begin{eqnarray}
	{\mathbf  q}(r) = \left( \begin{array}{cc}
	u(r) \\ x(r) \, \end{array} \right) .
\end{eqnarray}
Here, $x(r)$ is the logarithm of the energy dissipation rate
$\epsilon_{r}$ normalised by its (scale independent) expectation
value: $x(r) = \ln\left( \epsilon_{r}/\bar{\epsilon}\right)$.

The stochastic process underlying the evolution of ${\mathbf  q}(r)$
in the scale $r$ is Markovian, if the conditional pdf $p\left( \, 
{\mathbf  q}(r_{1}) \, | \, {\mathbf  q}(r_{2}), {\mathbf  q}(r_{3}),
...,{\mathbf  q}(r_{N}) \, \right)$ with $r_{1}\leq r_{2} \leq ...  \leq
r_{N}$ fulfills the relation:
\begin{equation}
	p\left( \, {\mathbf  q}(r_{1}) \, | \, {\mathbf  q}(r_{2}),
	{\mathbf  q}(r_{3}), ...  , {\mathbf  q}(r_{n}) \, ) = p( \,
	{\mathbf  q}(r_{1}) \, | \, {\mathbf  q}(r_{2}) \, \right) .
	\label{MarkovCond}
\end{equation}
$p\left( \, {\mathbf  q}(r_{1}) \, | \, {\mathbf  q}(r_{2}),{\mathbf
q}( r_{3}), ...  , {\mathbf  q}(r_{n}) \, \right)$ denotes the
probability for finding certain values for $u$ and $x$ at some scale
$r_{1}$, provided that the values of ${\mathbf  q}$ at all larger
scales $r_{2}, r_{3}, \ldots , r_{N}$ are known. The condition
(\ref{MarkovCond}) simply states that the transition from an eddy 
at scale $r_{2}$ characterized by $u(r_{2})$ and $x(r_{2})$ to the 
"state" ${\mathbf q}(r_{1})$ at scale $r_{1}$ should not depend on what
happens at larger scales.

If the conditional pdfs fulfill the Markov condition (\ref{MarkovCond}), 
any $N$--point distribution of ${\mathbf  q}$ can be expressed as a 
product of conditional pdfs:
\begin{eqnarray}
	p\left( \, {\mathbf q}(r_{1}),{\mathbf q}(r_{2}),..., {\mathbf
	q}(r_{N}) \, \right) & = & p \left( \, {\mathbf q}( r_{1}) \,
	| \, {\mathbf q}(r_{2}) \, \right) \times p\left( \, {\mathbf
	q}( r_{2}) \, | \, {\mathbf q}(r_{3}) \, \right) \times ... 
	\nonumber \\
	& & \times p\left( \, {\mathbf q}(r_{N-1} ) \, | \, {\mathbf
	q}( r_{n}) \, \right) \times p\left( \, {\mathbf q}(r_{N}) \,
	\right) .
	\label{chain}
\end{eqnarray}
Equation (\ref{chain}) is a remarkable statement: The knowledge of the
conditional pdf $p\left( \, {\mathbf q}(r) \, | \, {\mathbf
q}_{0}(r_{0}) \, \right) $ (for arbitrary scales $r$ and $r_{0}$ with
$r \leq r_{0}$) is sufficient to determine any $N$--point pdf of
${\mathbf q}$, i.e.: the entire information about the stochastic
process is encoded in the conditional pdf.

Furthermore, it is well-known that for Markov processes the evolution
of the conditional pdf in the scale $r$ can be described by the
Kramers-Moyal expansion, a partial differential equation for $p\left(
\, {\mathbf q}(r) \, | \, {\mathbf q}_{0}(r_{0}) \, \right)$ in the
variables ${\mathbf q}$ and $r$.  According to Pawula's theorem, this
expansion truncates after the second term if the fourth order
expansion coefficient vanishes.  In this case, the Kramers-Moyal
expansion reduces to the Fokker--Planck equation \footnote{Note that
we multiplied both sides of the Fokker-Planck equation with $r$ (in
contrast to the usual definition as, for example, given in
\cite{Risken}).  The factor $r$ on the right side of eq. 
(\ref{FoplaCond}) can be found in the definition (\ref{MkDef}) of the
conditional moments ${\mathbf M}^{(k)}$.  }:
\begin{eqnarray}
	- r \frac{\partial}{\partial r} p( \, {\mathbf  q},r \, | \,
	{\mathbf  q}_{0},r_{0} \, ) 
	& = & -  \sum\limits_{i=1}^{2}
	\frac{\partial}{\partial q_{i}} \left( \;
	D^{(1)}_{i}({\mathbf  q},r) \; p(\, {\mathbf  q},r \, |\,
	{\mathbf  q}_{0},r_{0}) \; \right) 
	\nonumber \\
	&& +  \sum\limits_{i,j=1}^{2} \frac{\partial^2}{\partial
	q_{i} \partial q_{j}} \left( \; D^{(2)}_{ij}
	({\mathbf  q},r) \; p( \, {\mathbf  q},r \, | \, {\mathbf
	q}_{0},r_{0} \, ) \; \right) .
	\label{FoplaCond}
\end{eqnarray}
Note that we changed notation writing $p( \, {\mathbf q},r \, | \,
{\mathbf q}_{0}, r_{0} \, )$ instead of $p\left( \, {\mathbf q}(r) \,
| \, {\mathbf q}_{0}(r_{0}) \, \right)$.  This notation is chosen in
order to indicate that the conditional pdf is a function of the scale
$r$ (although $r$, of course, is not a stochastic variable).

By multiplying the Fokker--Planck equation with $p({\mathbf
q}_{0},r_{0})$ and integrating with respect to ${\mathbf q}_{0}$, it
can be shown that the same equation also describes the $r$--evolution
of the pdf $p({\mathbf q},r)$. Mathematically, the drift vector
${\mathbf D}^{(1)}$ and the diffusion matrix ${\mathbf D}^{(2)}$ are
defined via the limit
\begin{eqnarray}
    D^{(1)}_{i}({\mathbf  q},r)  & = & \lim_{\Delta r \rightarrow 0} \,
    M^{(1)}_{i}({\mathbf  q},r, \Delta r) , \nonumber \\
    D^{(2)}_{ij}({\mathbf  q},r)  & = & \lim_{\Delta r \rightarrow 0} \,
    M^{(2)}_{ij}({\mathbf  q},r,\Delta r) , \label{DkDef}
\end{eqnarray}    
where the coefficients ${\mathbf M}^{(k)}$ are given by:
\begin{eqnarray}
	M^{(1)}_{i}({\mathbf q},r,\Delta r) & = & \frac{r}{\Delta r}
	\left< \; \left.  \left( \, q_{i}'(r-\Delta r) - q_{i}(r) \,
	\right) \, \right| \, {\mathbf q},r \; \right> \; , \nonumber
	\\
	M^{(2)}_{ij}({\mathbf  q},r,\Delta r) & = & \frac{r}{2 \Delta
	r} \left< \; \left( \, q_{i}'(r-\Delta r) - q_{i}(r)
	\right) \right.  \, \times 
	\nonumber \\
	&& \left.  \left.  \times \, \left( \, q_{j}'(r-\Delta r) -
	q_{j}(r) \, \right) \right| {\mathbf  q},r \; \right> . 
	\label{MkDef}
\end{eqnarray}
The coefficients ${\mathbf M}^{(k)}$ are nothing but conditional
expectation values of the veloctiy increment and the energy
dissipation rate, respectively, and can easily be determined from
experimental data.  One may therefore hope to find estimates for the
${\mathbf D}^{(k)}$ by extrapolating the measured conditional moments
${\mathbf M}^{(k)}$ towards $\Delta r = 0$ \cite{JFM}, see also 
\cite{Kantz, KantzReply}.

Alternatively, the stochastic process underlying the evolution of the 
variable ${\mathbf  q}$ in the scale $r$ can be described by the 
Langevin--equation, an ordinary stochastic differential equation for 
${\mathbf  q}(r)$:
\begin{eqnarray}
    - \frac{\partial}{\partial r} q_{i}(r) \; & = & \; f_{i}({\mathbf
    q},r) \, + \, \sum\limits_{j=1}^{2} g_{ij}({\mathbf q},r)
    \Gamma_{j}(r) \label{Langevin}
\end{eqnarray}
The components of the vector ${\mathbf \Gamma}(r)$ represent the
stochastic influences acting on the process.  It can be shown that a
variable which is described by eq.  (\ref{Langevin}) is Markovian, if
and only if the $\Gamma_{i}(r)$ are $\delta$--correlated stochastic
forces with zero mean.  If furthermore the pdf of the
stochastic forces are Gaussian, i.e. if ${\mathbf \Gamma}(r)$ is
$\delta$--correlated white noise, the Kramers-Moyal expansion stops
after the second term and the conditional pdf $p( \, {\mathbf
q},r|{\mathbf q}_{0},r_{0} \, )$ is described by the Fokker-Planck
equation.

In that case the functions ${\mathbf f}({\mathbf q},r)$ and ${\mathbf
g}({\mathbf q},r)$ can be calculated from the drift vector ${\mathbf
D}^{(1)}$ and the diffusion matrix ${\mathbf D}^{(2)}$.  In It\^{o}'s
formalism of stochastic calculus, ${\mathbf f}({\mathbf q},r)$ and
${\mathbf g}({\mathbf q},r)$ are given by
\begin{eqnarray}
	f_{i}({\mathbf q},r) & = & \frac{1}{r} D^{(1)}_{i}({\mathbf q},r) , 
	\nonumber \\
	g_{ij}({\mathbf q},r) & = & \left( \sqrt{ \frac{1}{r} {\mathbf
	D^{(2)}}} \right)_{ij} ,
	\label{FundGvonD12}
\end{eqnarray}
where $\sqrt{ {\mathbf D}^{(2)} }$ is to be calculated by diagonalizing the 
matrix ${\mathbf D}^{(2)}$, taking the square root of each element 
of the diagonalized matrix and transforming the result back into the 
original system of coordinates.

The Langevin--equation offers an alternativ way to check the Markovian
properties of a stochastic variable.  The idea which was originally
proposed in \cite{LangevinOriginalzitat} is to estimate the
coefficients ${\mathbf D}^{(1)}$ and ${\mathbf D}^{(2)}$ from
experimental data according to equations (\ref{MkDef}) and
(\ref{DkDef}) and to calculate the functions ${\mathbf f}$ and
${\mathbf g}$ according to equation (\ref{FundGvonD12}).  Having
determined ${\mathbf f}$ and ${\mathbf g}$ in that way, the
Langevin--equation (\ref{Langevin}) can be used to extract ${\mathbf
\Gamma}(r)$ from the (measured) derivatives of the $q_{i}(r)$.  If the
realizations of the stochastic force obtained by this method are
$\delta$--correlated with zero mean and a Gaussian distribution, the
process is governed by the Fokker-Planck equation.  We will use this
method here instead of the one proposed in \cite{JFM}, since for
multidimensional stochastic variables it is hardly possible to check
the Markov condition (\ref{MarkovCond}) directly by means of
multiconditional pdfs.  Also the numerical cost for the estimation of
the coefficients ${\mathbf D}^{(k)}$ of order three and higher grows
considerably with the order k (growth like $2^k$).

\section{The results of the one dimensional Markov analysis}\label{eindim}

Here we briefly recall the results of the one dimensional analysis for
the velocity increment and the energy dissipation rate, respectively. 
Detailled discussions of the results for the velocity increment can be
found in refs.  \cite{PRLTurb, PhysicaD, JFM, unpublished}, for recent
results on the Markov analysis of the energy dissipation rate we
refer to \cite{AntoineEnergie, AntoineLangevin}.

In the case of the one dimensional analysis of the velocity increment,
the Markov condition (\ref{MarkovCond}) was found to be valid for
scales $r_{i}$ and differences of scales $\Delta r=r_{i+1}-r_{i}$
larger than a certain scale $l_{mar}$, which is of the order of
magnitude of the Taylor microscale $\lambda$.  The turbulent cascade
thus exhibits an elementary step size.  This phenomenon may be seen in
analogy to the mean free path of molecules undergoing a Brownian
motion.

Within the range of scales for which the Markovian properties are
fulfilled, i.e. for $\Delta r > \lambda$, the conditional moments 
\begin{equation}
    M^{(k)}(u,r,\Delta r) = \frac{r}{ k!  \Delta r} \left< \left(
    u'(r-\Delta r) - u(r) \right)^k | u(r) \right>
\end{equation}
of the velocity increment show a linear dependence (with small second
order corrections) on $\Delta r$ and can thus be extrapolated towards
$\Delta r = 0$.  Furthermore, it can be shown that the fourth order
coefficient $D^{(4)}$ does not have an important influence on the
evolution of $p(u,r)$ and can be neglected \cite{JFM}.  The pdf of the
velocity increment is therefore governed by the Fokker--Planck
equation:
\begin{equation}
    - r \frac{\partial}{\partial r} p(u,r) = -
    \frac{\partial}{\partial u} \left( D^{(1)}(u,r) p(u,r) \right) +
    \frac{\partial^2}{\partial u^2} \left( D^{(2)}(u,r) p(u,r) 
    \right) \, .
    \label{FoplaEinDimInk}
\end{equation}
Drift and diffusion coefficient turn out to be linear and 
quadratic functions of $u$, respectively:
\begin{eqnarray}
    D^{(1)}(u,r) & = & - \gamma(r) u , \nonumber\\
    D^{(2)}(u,r) & = & \alpha(r) - \delta(r) u + \beta(r)u^2 .
    \label{D12EinDimInk}
\end{eqnarray}
When the scale $r$ is given in units of the Taylor microscale
$\lambda$, the linear term $\gamma$ of $D^{(1)}$ exhibits an universal
dependence on scale $r$, independent of the Reynolds number
\cite{unpublished,ChrisDiss}:
\begin{equation}
    \gamma(r) \approx \frac{2}{3} + 0.2 \sqrt{r/\lambda} . 
    \label{GammaEinDimInk}
\end{equation}
The coefficients $\alpha(r)$ and $\delta(r)$ are linear functions of
the scale $r$ with slopes which decrease with increasing Reynolds
number.  The quadratic term $\beta$ shows an only weak dependence on
$r$ but increases significantly with $Re$ \cite{unpublished,ChrisDiss}.

The analogous analysis for the energy dissipation rate
\cite{AntoineEnergie, AntoineLangevin} shows that the pdf $p(x,r)$ of
the logarithmic energy dissipation rate $x(r)$ is also governed by a
Fokker--Planck equation:
\begin{equation}
    - r \frac{\partial}{\partial r} p(x,r) = - 
    \frac{\partial}{\partial x} \left( D^{(1)}(x,r) p(x,r) \right) + 
    \frac{\partial^2}{\partial x^2} \left( D^{(2)}(x,r) p(x,r) 
    \right) . \label{FoplaEindimX}
\end{equation}
Again, the drift coefficient $D^{(1)}$ shows a linear dependence on 
its argument $x$, albeit with a positive slope and an additional 
constant term:
\begin{eqnarray}
    D^{(1)}(x,r) & = & F(r) + G(r) x . \label{D1EindimX}
\end{eqnarray}
In \cite{AntoineEnergie, AntoineLangevin} the scale dependence of the
coefficients $F$ and $G$ was expressed by means of the invers
logarithmic scale $l = \ln \left( L/r \right)$.  Rewritten in terms of
the linear scale $r$, the parametrizations given in
\cite{AntoineEnergie, AntoineLangevin} read:
\begin{eqnarray}
    F(r) & = & - A_{0} \left( \frac{r}{L} \right)^{-A_{1}} , \nonumber \\
    G(r) & = & B_{0} + B_{1} \ln \left( \frac{r}{L} \right) . 
    \label{FundGEindim}
\end{eqnarray}
The diffusion coefficient $D^{(2)}(x,r)$ can in a first order
approximation taken to be constant in $x$ \cite{AntoineEnergie} but is
found to depend on the scale $r$ \cite{AntoineLangevin}:
\begin{eqnarray}
    D^{(2)}(x,r) \approx D(r) = C_{0} \left( \frac{r}{L}
    \right)^{-C_{1}} .  \label{D2EindimX}
\end{eqnarray}
With a diffusion coefficient that does not depend on $x$, the
solutions of the Fokker--Planck equation (\ref{FoplaEindimX}) are
Gaussian \cite{Risken}, i.e. the pdf of the averaged energy
dissipation rate is lognormal in agreement with Kolmogorov's
\cite{K62} assumption.  But it also follows from eqs. 
(\ref{FoplaEindimX}), (\ref{D1EindimX}), (\ref{FundGEindim}) and
(\ref{D2EindimX}) that the standard deviation of this pdf is not
described by a logarithmic dependence as assumed by Kolmogorov to
provide scaling behaviour \cite{AntoineLangevin}.  It can also be seen
from experimental data that the constant value for $D^{(2)}(x,r)$
according to eq.  (\ref{D2EindimX}) is a first order approximation;
the data presented in \cite{AntoineEnergie} (fig. 2) reveal a weak
but nonetheless significant dependence on $x$.

\section{The experimental setup}\label{setup}

The data set used for the analysis consists of $10^8$ samples of the
local velocity measured in a cryogenic axisymmetric helium gas jet at
a Reynolds number of $115000$.  The measurement was done in the center
of the jet at a vertical distance of $40 D$ ($D=2mm$ is the diameter
of the nozzle) from the nozzle using a specially adapted hotwire
anemometer with a spatial resolution of $3.4 \mu m$ \cite{ChanalDiss,
ChanalPaper}.  We use Taylor's hypothesis of frozen turbulence to
convert time lags into spatial displacements.  With the sampling
frequency of $91.9 kHz$ and a mean velocity of $0.62 m/s$, the spatial
resolution of the measurement is $6.8 \mu m$.  Following the
convention chosen in \cite{JFM}, velocities and velocity increments
are given furtheron in units of $\sigma_{\infty}$.  It is defined as
$\sigma_{\infty} = \sqrt{2}\sigma$, where $\sigma$ is the standard
deviation of velocity fluctuations.  For the data set under
considerations, $\sigma_{\infty}= 0.2 m/s$.

For the integral length scale $L$ we obtain a value of $3.5 mm$, the
Taylor microscale $\lambda$ is $120 \mu m$ and the dissipation scale
$\eta$ was estimated to be approximately $6 \mu m$.  For further
details on the experimental setup we refer the reader to
\cite{ChanalDiss,ChanalPaper}.

\begin{figure}[ht]
  \begin{center}
    \epsfig{file=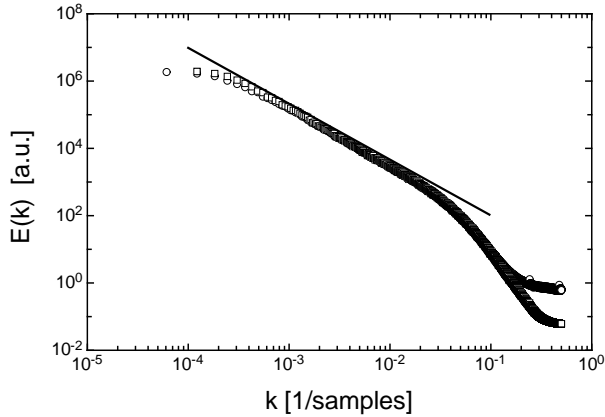, width=8.0cm}
  \end{center}
 \caption{ \it The wave number spectra $E(k)$ of the original data (circles) 
 and the smoothed data set (squares). The wave number $k$ is given in 
 units of samples. The straight line indicates scaling behaviour 
 according to K41: $E(k) \propto k^{-5/3}$.
 }
 \label{powerplot}
\end{figure}

The energy dissipation rate is estimated by its one--dimensional
surrogate \cite{turb}
\begin{equation}
    \epsilon(x) = \frac{15}{\nu} \left( \frac{\partial v(x)}{\partial
    x} \right)^{2} .
\end{equation}
For experimental reasons the derivative of the velocity field has
to be approximated by a finite difference:
\begin{equation}
    \frac{\partial v(x)}{\partial x} \approx \frac{v(x+\Delta) - 
    v(x)}{\Delta} .
\end{equation}
As the dissipative scale $\eta$ is resolved by the measurement (the
resolution of the sensor is $3,4 \mu m$, while $\eta$ is approximately
$6 \mu m$), it is in principle allowed to set $\Delta = 1$ (in units
of samples).  However, from the wave number spectrum of the data shown
in figure \ref{powerplot}, it becomes evident that the measurement is 
dominated by white noise for small scales (below $\Delta \approx 5$),
which would lead to incorrect results for the energy dissipation rate if
$\Delta=1$ was used to estimate $\partial v / \partial x$.
%
%

We therefore applied a digital low pass filter to the data
multiplying the Fourier coefficients with the spectral filter function
\begin{equation}
    \phi(k) = \frac{1}{1+(k/k_{0})^4} .
\end{equation}
The cutoff wave number was chosen to be $k_{0}=0.2$, according to the
observed transition to white noise for wave numbers $k>0.2$ (see fig. 
\ref{powerplot} which also displays the power spectrum of the filtered
data set). Details on the method can be found in \cite{numrec}. 

A digital filter is of course a serious manipulation of the data and
it is by no means obvious that the smoothed data represent the "real"
velocity signal in a better approximation than the original data set. 
Therefore we compared the various results obtained from the original
and the smoothed data.  In both cases we used $\Delta = 5$ for the
estimation of the energy dissipation rate.  It is found that for both
cases the coefficients ${\mathbf D}^{(k)}$ show the same functional
dependencies on their arguments $u$, $x$ and $r$.  Thus the
coefficients calculated from the original and smoothed data set,
respectively, are identical up to constant factor \cite{ChrisDiss}. 
For the purpose of the analysis presented in this paper, these effects
are of no importance.  We therefore restrict the discussion to the
results obtained from the smoothed signal.  For the filtered data set,
the mean energy dissipation rate $\bar{\epsilon}$ is $0.52 m^2 / s^3$.

\section{Experimental results}\label{results}

\begin{figure}[ht]
  \begin{center}
    \epsfig{file=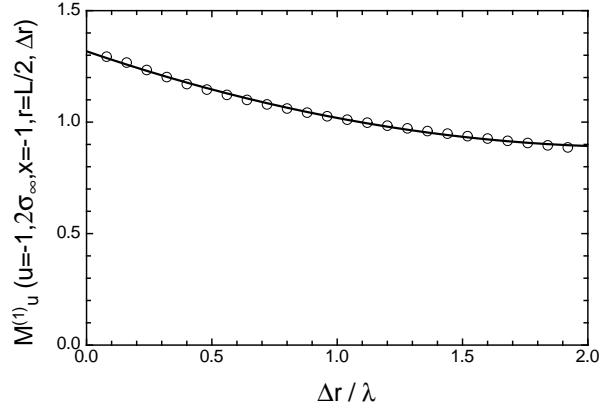, width=8.0cm}
  \end{center}
 \caption{ \it The coefficient $M^{(1)}_{u}(u,x,r,\Delta r)$ at scale
 $r=L/2$ as a function of $\Delta r$ for $u=+\sigma_{\infty}$ and
 $x=+1$ (circles). The data can be described and extrapolated by a 
 polynomial of degree two in $\Delta r$ (line).
 }
 \label{M1uVonDeltaR}
\end{figure}

Let us start with the $u$--component of the driftvector. To this end, 
we have to calculate the conditional moment 
\begin{equation}
    M^{(1)}_{u}(u,x,r,\Delta r) \; = \; \frac{r}{\Delta r} \left< \,
    u'(r-\Delta r) - u(r) \, | \, u(r),x(r) \, \right> \label{M1uDef}
\end{equation}
for various values of $u$, $x$, $r$ and $\Delta r$ and try to
extrapolate it towards $\Delta r = 0$ according to equation
(\ref{DkDef}).  Figure \ref{M1uVonDeltaR} shows the coefficient
$M^{(1)}_{u}$ at scale $r=L/2$ for $u=+\sigma_{\infty}$ and $x=+1$ as
a function of $\Delta r$.  Over the whole range of (differences of)
scales $0 < \Delta r \leq 2 \lambda$, the dependence of $M^{(1)}_{u}$
on $\Delta r$ can in good approximation be described by a polynomial
of degree two in $\Delta r$ (see \cite{Kantz, KantzReply}), thus
allowing for an extrapolation towards $\Delta r = 0$.
%
%

\begin{figure}[ht]
  \begin{center}
    \epsfig{file=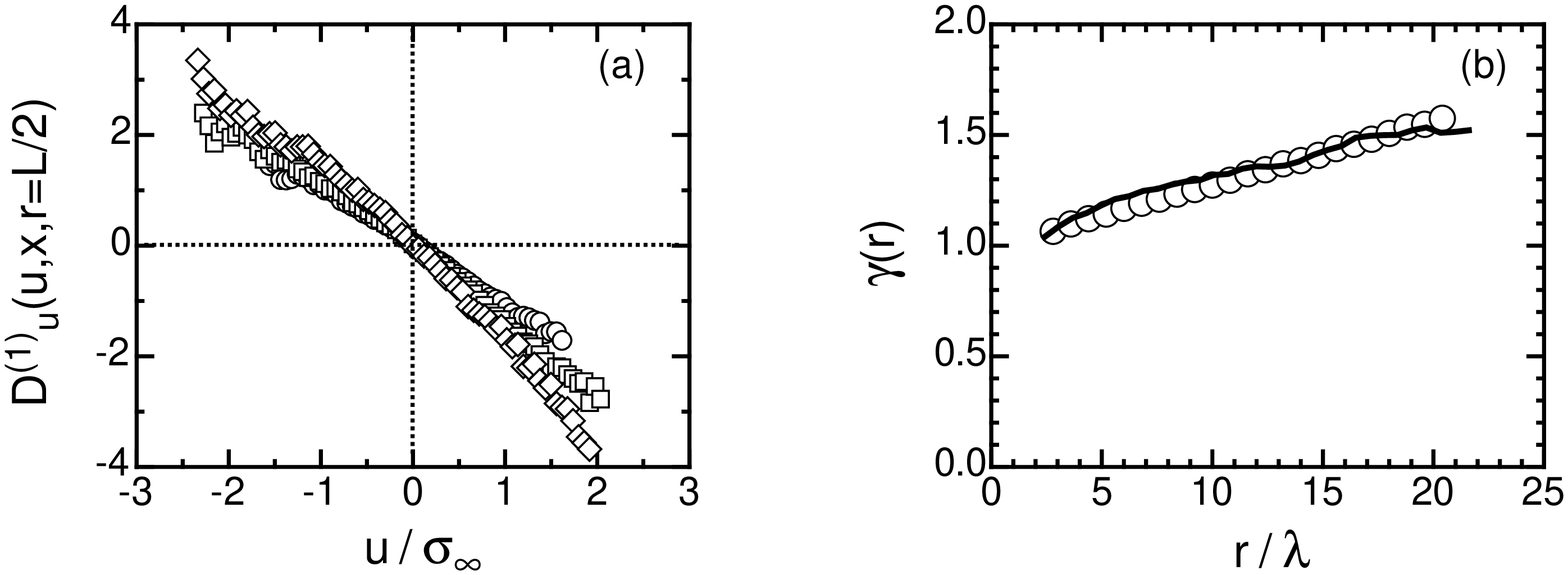, width=12.0cm}
  \end{center}
 \caption{ \it (a): The $u$--component $D^{(1)}_{u}(u,x,r)$ of the
 drift vector at scale $r=L/2$ as a function of the velocity increment
 $u$ for $x=-1$ (circles), $x=0$ (squares) and $x=+1$ (diamonds). 
 \newline (b): The slope $\tilde{\gamma}(r)$ of $D^{(1)}_{u}(u,r)$ as a function of
 the scale $r$ (circles).  $\tilde{\gamma}(r)$ turns out to be identical with the
 slope $\gamma(r)$ of the one dimensional drift coefficient (line).  }
 \label{D1u}
\end{figure}

Figure \ref{D1u}(a) shows the result of the extrapolation for
$D^{(1)}_{u}(u,x,r)$ at scale $r=L/2$ for various values of $x$ as a
function of the velocity increment $u$. The $u$--component of the
drift vector shows a linear dependence on the velocity increment and
only weak variations for different values of $x$. To a first order
approximation, the dependence of $D^{(1)}_{u}$ on $x$ can thus be neglected
and we obtain:
\begin{eqnarray}
    D^{(1)}_{u}(u,x,r) \approx D^{(1)}_{u}(u,r) = -  \tilde{\gamma}(r)u . 
    \label{D1uZweiDim}
\end{eqnarray}
%
%

The slope $\tilde{\gamma}(r)$, which for a given scale $r$ is obtained
by averaging the results of the fits (\ref{D1uZweiDim}) for different
$x$, has a value of $1.1$ at the scale $r=L/2$.  By performing this
procedure at several scales $r$ we are able to specify the scale
dependence of $\tilde{\gamma}(r)$, see fig.  \ref{D1u}(b).  It turns
out that the function $\tilde{\gamma}(r)$ is identical with the slope
$\gamma(r)$ of the drift coefficient of the one dimensional
Fokker--Planck equation (\ref{D12EinDimInk}) for $p(u,r)$.

The finding that $\tilde{\gamma}(r)$ is identical with the one
dimensional coefficient $\gamma(r)$ may be surprising at first sight,
but is a direct consequence of the fact that $D^{(1)}_{u}$ does
(approximately) not depend on $x$.  This can be seen by considering
the Fokker--Planck equation (\ref{FoplaCond}) for the joint pdf
$p(u,x,r)$:
\begin{eqnarray}
    - r \frac{\partial}{\partial r} p(u,x,r) & = & - 
    \frac{\partial}{\partial u} \left( D^{(1)}_{u} p \right) - 
    \frac{\partial}{\partial x} \left( D^{(1)}_{x} p \right) 
    \nonumber \\ 
    && + \frac{\partial^2}{\partial u^2} \left( D^{(2)}_{uu} p \right) + 
    \frac{\partial^2}{\partial x^2} \left( D^{(2)}_{xx} p \right) + 2 
    \frac{\partial^2}{\partial u \partial x} \left( D^{(2)}_{ux} p \right)
    \label{Fopla2dimKomp}
\end{eqnarray}
From the two dimensional equation (\ref{Fopla2dimKomp}) the equation
for $p(u,r)$ can be derived by integrating with respect to $x$:
$p(u,r) = \int p(u,x,r) dx$.  Assuming that the product
$D^{(k)}_{i}(u,x,r) p(u,x,r)$ vanishes for $x \rightarrow \pm \infty$
one obtains:
\begin{eqnarray}
    - r \frac{\partial}{\partial r} p(u,r) & = & - r \frac{\partial}
    {\partial r} \int p(u,x,r) dx 
    \nonumber \\ 
    & = & - \frac{\partial}{\partial u} \left( D^{(1)}_{u}(u,r) \int
    p(u,x,r) dx \right) \nonumber \\
    && + \frac{\partial^2}{\partial u^2} \left( \int
    D^{(2)}_{uu}(u,x,r) p(u,x,r) dx \right) 
    \nonumber \\
    & = & - \frac{\partial}{\partial u} \left( D^{(1)}(u,r) p(u,r) 
    \right) \nonumber \\
    && +  \frac{\partial^2}{\partial u^2}  \underbrace{ \int
    D^{(2)}_{uu}(u,x,r) p(x,r|u,r) dx}_{D^{(2)}(u,r)} \, p(u,r)  .
    \label{EindimAusZweidim}
\end{eqnarray}
The result of this calculation is a one dimensional Fokker--Planck
equation for $p(u,r)$ with a drift coefficient $D^{(1)}(u,r)$ that is
identical to its two dimensional counterpart $D^{(1)}_{u}(u,r)$.

\begin{figure}[ht]
  \begin{center}
    \epsfig{file=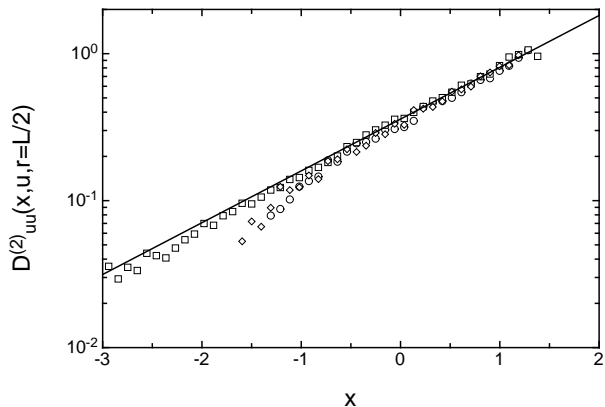, width=8.0cm}
  \end{center}
 \caption{ \it The $uu$--component $D^{(2)}_{uu}(u,x,r)$ of the
 diffusion matrix at scale $r=L/2$ as a function of $x$ for
 $u=-\sigma_{\infty}$ (circles), $u=0$ (squares) and
 $u=+\sigma_{\infty}$ (diamonds).  The coefficient does not depend on
 $u$ and can be described by an exponential in $x$ (the straight line
 represents a fit to the data for $u=0$ according to eq. 
 (\ref{D2uuFunk})).  }
 \label{D2uuVonX}
\end{figure}

The second order coefficient $D^{(2)}_{uu}(u,x,r)$ can be estimated by
an extrapolation of the conditional second order moments
\begin{equation}
    M^{(2)}_{uu}(u,x,r,\Delta r) = \frac{r}{2 \Delta r} \left< \left(
    u'(r-\Delta r) - u(r) \right)^2 | u(r),x(r) \right>
\end{equation}
towards $\Delta r \rightarrow 0$ in the same way as described above 
for the first order coefficient $D^{(1)}_{u}$. As can be seen in figure
\ref{D2uuVonX}, the coefficient $D^{(2)}_{uu}$ does not depend on the
velocity increment $u$ and shows an exponential dependence on $x$:
\begin{eqnarray}
    D^{(2)}_{uu}(u,x,r) = D^{(2)}_{uu}(x,r) = a_{0}(r) \exp \left( 
    a_{1}(r) x \right) \label{D2uuFunk}
\end{eqnarray}

%
%

\begin{figure}[ht]
  \begin{center}
    \epsfig{file=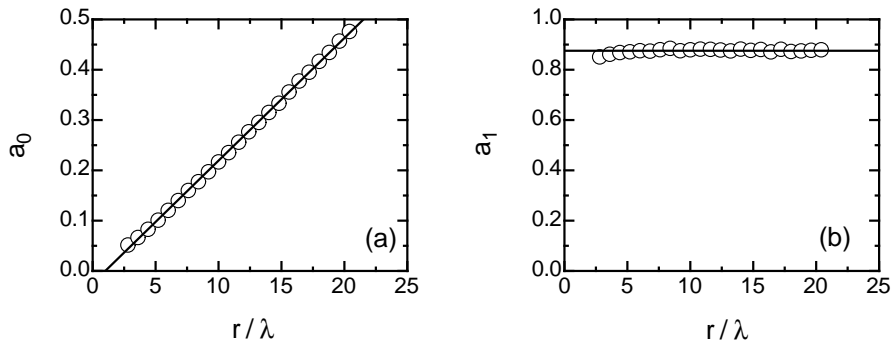, width=12.0cm}
  \end{center}
 \caption{ \it The coefficients $a_{0}$ (a) and $a_{1}$ (b) defined in 
 eq. (\ref{D2uuVonX}) as functions of the scale $r$. $a_{0}$ is a 
 linear function of $r$, $a_{1}$ is approximately constant (straight 
 lines).
 }
 \label{D2uuKoeffs}
\end{figure}

The coefficients $a_{0}$ and $a_{1}$ defined in eq.  (\ref{D2uuFunk})
show simple dependencies on the scale $r$ (see figure
\ref{D2uuKoeffs}): $a_{0}$ is a linear function of $r$, while $a_{1}$
is constant:
\begin{eqnarray}
    a_{0}(r) & \approx & 0.02 \left( \frac{r}{\lambda} - 1 \right) , 
    \nonumber \\
    a_{1} & \approx & 0.9 . \label{D2uuValues}
\end{eqnarray}
Note that the linear dependence of $a_{0}$ on the scale $r$ as
specified in eq.  (\ref{D2uuValues}) leads to negative values for
$D^{(2)}_{uu}$ on scales smaller than the Taylor microscale $\lambda$. 
This is in contradiction to the definitions (\ref{DkDef}) and
(\ref{MkDef}) of the diffusion matrix since, according to those
definitions, the diagonal--elements of ${\mathbf D}^{(2)}$ are
positive quantities.  The finding of negative values for
$D^{(2)}_{uu}$ on scales smaller than $\lambda$ indicates that,
similar to the one dimensional case, the Markovian property cannot be
fulfilled for such small scales.

%
%

The fact that $D^{(2)}_{uu}(u,x,r)$ does not depend on $u$ is a very
interesting result since it states that the effect of intermittency is
caused by the stochastic nature of the energy dissipation rate.  This
can be seen by considering equation (\ref{EindimAusZweidim}) for the
(hypothetical) K41--case that $\epsilon_{r}$ is not a stochastic
variable but constant.  In this case the conditional pdf $p(x,r|u,r)$
in equation (\ref{EindimAusZweidim}) is formally given by $p(x,r|u,r)
= p(x,r) = \delta(x-1)$ and the one dimensional coefficient $D^{(2)}$
can easily be calculated: $D^{(2)}= \int D^{(2)}_{uu}(x,r) p(x,r|u,r)
dx = D^{(2)}_{uu}(x=1,r)=D^{(2)}(r)$.  With a diffusion coefficient
that does not depend on $u$ the resulting Fokker--Planck equation
(\ref{EindimAusZweidim}) for $p(u,r | \epsilon_{r}=const)$ can be
shown to be solved by a Gaussian distribution \cite{Risken}.  The
effect of intermittecy can thus be clearly traced back to the
statistics of $\epsilon_{r}$.  This result is in full agreement with
the results for the conditional pdf $p(u(r)|\epsilon_{r})$ presented
in \cite{Gagne, AntoineBedingt}.

\begin{figure}[ht]
  \begin{center}
    \epsfig{file=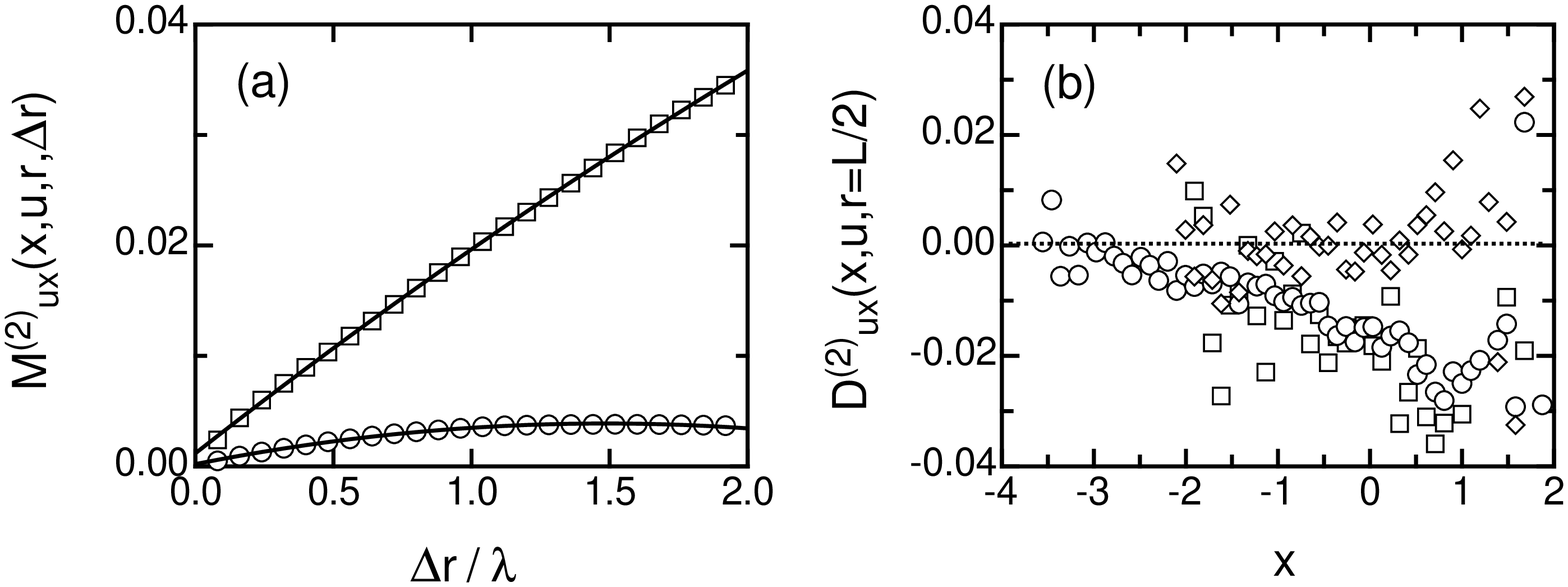, width=12.0cm}
  \end{center}
 \caption{ \it (a): The conditional moment $M^{(2)}_{ux}(u,x,r,\Delta 
 r)$ at scale $r=L/2$ as a function of $\Delta r$ for $u=-0.6 
 \sigma_{\infty}, x = -3$ (squares) and $u=+ 0.6 \sigma_{\infty}, 
 x=+1$ (circles). The data can be extrapolated by polynomials of 
 degree two (lines). \newline
 (b): The extrapolated coefficient $D^{(2)}_{ux}(u,x,r)$ at scale 
 $r=L/2$ as a function of $x$ for $u=-\sigma_{\infty}$ (squares), 
 $u=0$ (circels) and $u= + \sigma_{\infty}$ (diamonds).
 }
 \label{D2ux}
\end{figure}

Before proceeding with the $x$--components of drift vector and 
diffusion matrix, let us draw attention to the mixed term 
$D^{(2)}_{ux}$ of the diffusion matrix. This term can be estimated by 
extrapolating the conditional moments
\begin{equation}
    M^{(2)}_{ux} = \frac{r}{2 \Delta r} \left< \, \left( u'(r-\Delta 
    r) - u(r) \right) \left( x'(r-\Delta r) - x(r) \right) \, | \, 
    u(r),x(r) \, \right>
\end{equation}
towards $\Delta r \rightarrow 0$. Figure \ref{D2ux}(a) shows
the coefficient $M^{(2)}_{ux}$ at scale $r=L/2$ for exemplarily chosen
values of $u$ and $x$ as a function of $\Delta r$. The coefficient
exhibits large variations with $\Delta r$ and its absolute value shows a
strong decrease as $\Delta r$ goes to zero. This may be taken as a
first hint that $M^{(2)}_{ux}$ vanishes in the limit $\Delta r
\rightarrow 0$.
%
%

When the extrapolation towards $\Delta r \rightarrow 0$ is performed
by fitting polynomials of degree two to the data in the interval $0 <
\Delta r \leq 2 \lambda$ (see fig.  \ref{D2ux}a), we obtain values for
$D^{(2)}_{ux}$ which are small compared to the corresponding values of
$M^{(2)}_{ux}$ at finite $\Delta r$.  Furthermore, the coefficient
$D^{(2)}_{ux}$ shows fluctuations which are of the order of magnitude
of the values themselves (see fig.  \ref{D2ux}b). We take this as
evidence that the mixed coefficient $D^{(2)}_{ux}$ vanishes.

Given that the off--diagonal element $D^{(2)}_{ux}$ of the diffusion
matrix vanishes, the functions ${\mathbf g}(u,x,r)$ in the
Langevin--equation (\ref{Langevin}) can easily be calculated according
to equation (\ref{FundGvonD12}).  In It\^{o}'s formalism, ${\mathbf
g}$ is then simply given by:
\begin{eqnarray}
    g_{uu} = \sqrt{ \frac{1}{r} D^{(2)}_{uu} },  \qquad
    g_{ux} = g_{xu} = 0, \qquad 
    g_{xx} = \sqrt{ \frac{1}{r} D^{(2)}_{xx} } . \label{gAusD2Simpel}
\end{eqnarray}
With the results obtained so far the Langevin--equation 
(\ref{Langevin}) takes the form:
\begin{eqnarray}
    - \frac{\partial}{\partial r} u(r) & \; = \; & \frac{\gamma(r)}{r}
    u \, + \, \sqrt{ \frac{a_{0}(r)}{r} \exp\left( a_{1} x(r) \right)}
    \, \Gamma_{u}(r) , 
    \nonumber \\
    - \frac{\partial}{\partial r} x(r) & \; = \; & \frac{1}{r}
    D^{(1)}_{x}(u,x,r) \, + \, \sqrt{ \frac{1}{r} 
    D^{(2)}_{xx}(u,x,r)} \, \Gamma_{x}(r) .
    \label{LangevinII}
\end{eqnarray}
Even though the formulation (\ref{LangevinII}) of the twodimensional
Langevin--equation is yet incomplete, it can already be used to
extract the $u$--component $\Gamma_{u}(r)$ of the stochastic force
from measured realizations of $u(r)$, $\frac{\partial}{\partial r}
u(r)$ and $x(r)$.  These calculations are faciliated by the fact that
the diffusion coefficient $D^{(2)}_{uu}$ does not depend on the
velocity increment $u$.  It is therefore not necessary to distinguish
between the various definitions of stochastic calculus (It\^{o} and
Stratonovich, respectively), and a simple Euler--scheme can be applied
to calculate $\Gamma_{u}(r)$ (for details on the numerical scheme see,
for example, \cite{AntoineLangevin}).

\begin{figure}[ht]
  \begin{center}
    \epsfig{file=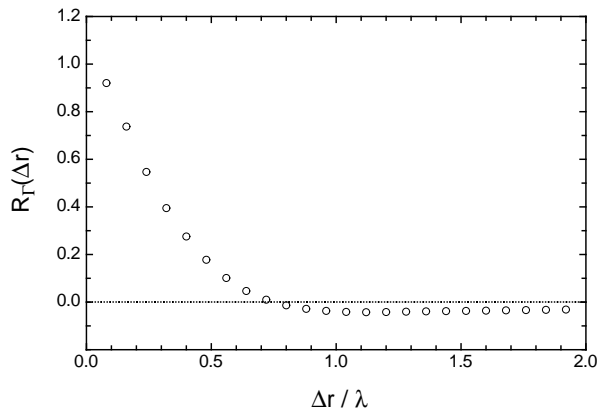, width=8.0cm}
  \end{center}
 \caption{ \it The autocorrelation function $R_{\Gamma}(\Delta r)$ of
 the $u$--component $\Gamma_{u}(r)$ of the stochastic force ${\mathbf
 \Gamma}(r)$ (open circles).  $R_{\Gamma}(\Delta r)$ exhibits finite
 values up to scales $\Delta r \approx \lambda$.  
 }
 \label{GammaKorr}
\end{figure}

The stoachstic process governing the $r$--evolution of the stochastic
variable ${\mathbf  q}(r)$ is Markovian, if the stochastic force
${\mathbf \Gamma}$ is $\delta$--correlated.  However, as can be seen
in figure \ref{GammaKorr}, the autocorrelation function
$R_{\Gamma}(\Delta r)$ of the stochastic force $\Gamma_{u}(r)$
exhibits nonzero values up to $\Delta r \approx \lambda$.  This
clearly indicates that the Markovian properties are fulfilled for
scales $r$ (and differences of scales $\Delta r$) larger than the
Taylor microscale $\lambda$ only.  Again, we recall that this result
is in agreement with the one dimensional analysis of the velocity
increment (see \cite{JFM}).

%
%

\begin{figure}[ht]
  \begin{center}
    \epsfig{file=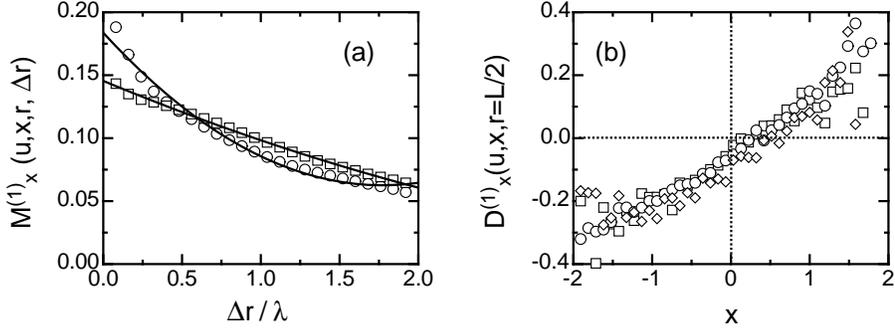, width=12.0cm}
  \end{center}
 \caption{ \it (a): The conditional moment $M^{(1)}_{x}(u,x,r,\Delta r)$ at 
 scale $r=L/2$ as a function of $\Delta r$ for $u=-0.6 
 \sigma_{\infty}, x = +1$ (circles) and $u=+ 0,6 \sigma_{\infty}, 
 x=+1$ (squares). The data can be extrapolated by polynomials of 
 degree two (lines). \newline
 (b): The extrapolated coefficient $D^{(1)}_{x}(u,x,r)$ at 
 scale $r=L/2$ as a function of $x$ for $u=-\sigma_{\infty}$ 
 (squares), $u=0$ (circles) and $u=+\sigma_{\infty}$ (diamonds).
 }
 \label{D1x}
\end{figure}

To complete the description of the turbulent cascade by the Markov
analysis, we still have to determine the coefficients $D^{(1)}_{x}$
and $D^{(2)}_{xx}$ by extrapolating the conditional moments
$M^{(1)}_{x}(u,x,r,\Delta r)$ and $M^{(2)}_{xx}(u,x,r,\Delta r)$,
respectively.  As shown in fig.  \ref{D1x}(a) for exemplarily chosen
values of $u$, $x$ and $r$, the first order coefficient $M^{(1)}_{x}$
can be described by a polynomial of degree two in $\Delta r$ over the
whole range of scales $0 < \Delta r \leq 2 \lambda$, which again
allows for an extrapolation towards $\Delta r = 0$.
%
%

The coefficient $D^{(1)}_{x}$ turns out to be a linear function of the 
logarithmic energy dissipation rate $x(r)$ and does not depend on the 
velocity increment $u$, see fig. \ref{D1x}(b):
\begin{eqnarray}
    D^{(1)}_{x}(u,x,r) = D^{(1)}_{x}(x,r) = F(r) + G(r) x 
    \label{D1xVonXFunk}
\end{eqnarray}
Since the coefficient $D^{(1)}_{x}$ does not depend on $u$, it has to
be identical with the coefficient $D^{(1)}$ of the one dimensional
Fokker--Planck equation (\ref{FoplaEindimX}) for $p(x,r)$.  This
follows from an analogous consideration leading to equation
(\ref{EindimAusZweidim}). 

\begin{figure}[ht]
  \begin{center}
    \epsfig{file=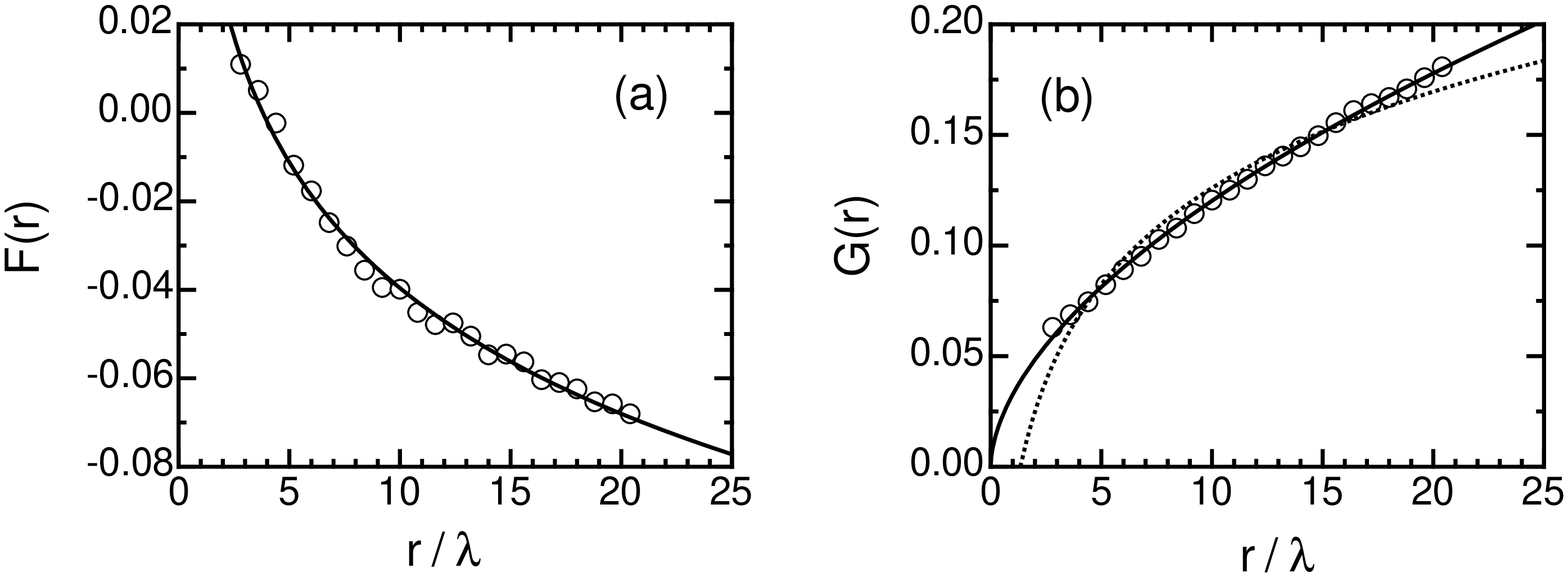, width=12.0cm}
  \end{center}
 \caption{ \it (a): The constant term $F(r)$ of the drift 
 coefficient $D^{(1)}_{x}$ as a function of the scale $r$. The full 
 line indicates a fit according to eq. (\ref{FundGZweiDim}). \newline
 (b): The slope $G(r)$ of $D^{(1)}_{x}$ as a function of $r$. Full 
 line: fit according to eq. (\ref{FundGZweiDim}); dotted line: fit 
 according to eq. (\ref{FundGEindim}).
 }
 \label{FundG}
\end{figure}

Plotting $F$ and $G$ as functions of the linear scale $r$, we find
that their scale dependence is best described by (see fig. 
\ref{FundG}):
\begin{eqnarray}
    F(r) & \approx & 0.05 - 0.04 \ln \left( \frac{r}{\lambda} 
    \right) \nonumber \\
    G(r) & \approx & 0.03 \left( \frac{r}{\lambda} 
    \right)^{0.57}. \label{FundGZweiDim}
\end{eqnarray}
Note that these parametrizations differ from those given for the
one dimensional coefficients in equation (\ref{FundGEindim}).  However,
the discrepancy between the results given in (\ref{FundGZweiDim}) and
(\ref{FundGEindim}) must not be taken too serious.  In
\cite{AntoineLangevin} $F$ and $G$ were plotted in terms of the invers
logarithmic length scale $l = \ln\left( L/r \right)$, which may suggest
a different functional dependence of those coefficients on the scale
than it is found here.  Furthermore, a data set at $R_{\lambda}=341$
was used, whereas for the data set used here the Taylor--Reynolds
number is $463$. In addition, as mentioned above, numerical values
such as those for $F(r)$ and $G(r)$ strongly depend on the method
chosen to estimate the derivative $\frac{\partial v}{\partial x}$ (see
chapter \ref{setup}).

%
%
%

%

\begin{figure}[ht]
  \begin{center}
    \epsfig{file=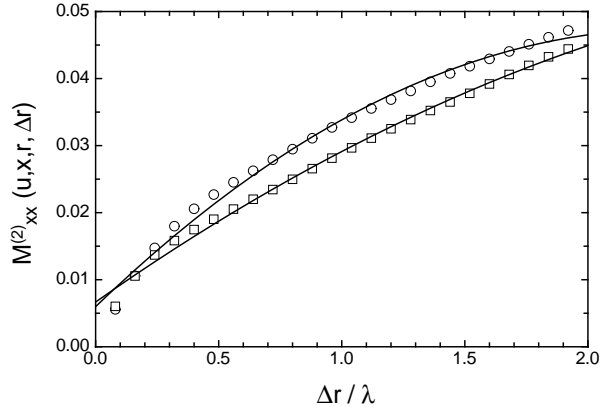, width=8.0cm}
  \end{center}
 \caption{ \it The conditional moment $M^{(2)}_{xx}(u,x,r,\Delta r)$ 
 at scale $r=L/2$ as a function of $\Delta r$ for $u=-0.6 
 \sigma_{\infty}, x = +1$ (circles) and $u=+ 0,6 \sigma_{\infty}, 
 x=+1$ (squares). The data were extrapolated by polynomials of 
 degree two (lines). 
 }
 \label{M2xxVonDeltaR}
\end{figure}

\begin{figure}[ht]
  \begin{center}
    \epsfig{file=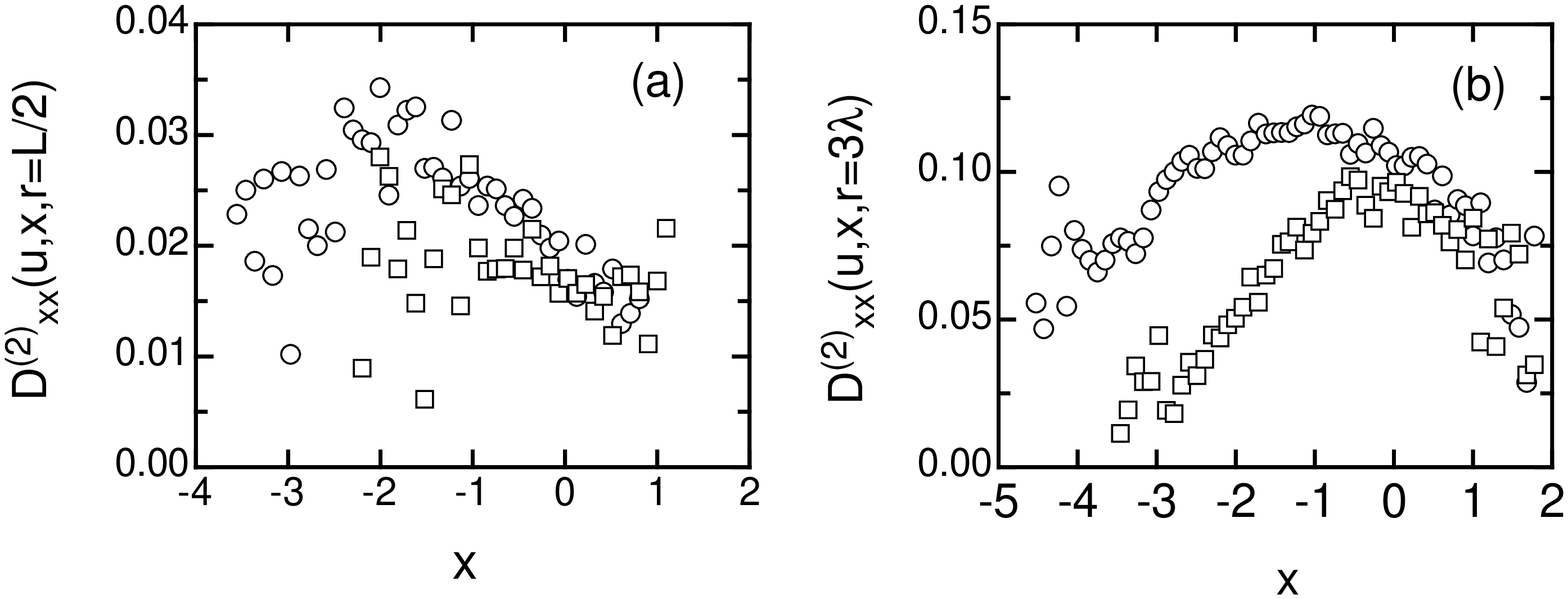, width=12.0cm}
  \end{center}
 \caption{ \it The $xx$--component $D^{(2)}_{xx}(u,x,r)$ of the 
 diffusion matrix as a function of the logarithmic dissipation rate 
 $x$. (a): At scale $r=L/2$ for $u=0$ (circles) and 
 $u=+\sigma_{\infty}$ (squares). (b): At scale $r=3 \lambda$ for $u=0$ 
 (circles) and $u=\frac{1}{4}\sigma_{\infty}$ (squares).
 }
 \label{D2xx}
\end{figure}

The last coefficient which remains to be calculated is the
$xx$--component of the diffusion matrix. Unfortunately, the
estimation of $D^{(2)}_{xx}(u,x,r)$ from the conditional moment 
\begin{equation}
    M^{(2)}_{xx}(u,x,r,\Delta r) = \frac{r}{2 \Delta r} \left< \left(
    x'(r-\Delta r) - x(r) \right)^2 | u(r),x(r) \right>
\end{equation}
yields several difficulties.  When plotted as a function of $\Delta
r$, the conditional moment $M^{(2)}_{xx}$ shows a strong decrease as
$\Delta r$ goes to zero (see fig. \ref{M2xxVonDeltaR}): The
extrapolated values of $D^{(2)}_{xx}$ are small compared to the values
of $M^{(2)}_{xx}$ for finite values of $\Delta r$.
%
%
Accordingly, the coefficient $D^{(2)}_{xx}$ at scale $r=L/2$ does not
exhibit systematic dependencies on its arguments $x$ or $u$, see fig. 
\ref{D2xx}(a).  Although the values seem to decrease for large values
of $|x|$ with a maximum at $x \approx -2$, the considerable scatter of
the data also allows to assume a constant value for $D^{(2)}_{xx}$.  A
different result is obtained for smaller scales, as shown exemplarily
for $r=4\lambda$ in fig.  \ref{D2xx}(b).  In this case the coefficient
$D^{(2)}_{xx}$ clearly exhibits a dependence on the energy dissipation
rate $x$ as well as a dependence on the velocity increment $u$. 
However, from the data presented in fig.  \ref{D2xx}(b) it is still
impossible to decide how the dependence of $D^{(2)}_{xx}$ is to be
parametrized; the data can be fitted with several functions (Gaussian
as well as Lorentzian distributions, for example) with almost equal
accuracy.  The question of whether one of these functions is to be
preferred requires further detailled experimental as well as
theoretical investigations and is the subject of an ongoing study. 


%
%

\section{Conclusion and Comments}\label{discussion}

Although further investigations will be necessary to complete the
two dimensional Markov analysis of the turbulent cascade, the results
obtained so far already allow for several interesting statements on
the joint statistical properties of the longitudinal velocity
increment and the averaged energy dissipation rate.

A particularly remarkable result is obtained for the $u$--component
$D^{(2)}_{uu}$ of the diffusion matrix, which is found not to depend on
the velocity increment.  This means that if the averaged energy
dissipation rate $\epsilon_{r}$ was not a stochastic variable but
constant, the pdf of $u(r)$ would reduce to a rather simple Gaussian
distribution.  The effect of small scale intermittency can thus
clearly be traced back to the stochastic nature of
$\epsilon_{r}$.  

So far, our results are in accordance with earlier experimental
investigations \cite{Gagne, AntoineBedingt} as well as with the
assumptions underlying Kolmogorov's models.  Significant new
information, however, are found for the $u$--component of the drift
vector.  We found that $D^{(1)}_{u}(u,x,r)$ does not depend on the
energy dissipation rate, in a first order approximation.  This means
that this coefficient is identical with the drift term of the one
dimensional Fokker--Planck equation for $p(u,r)$, i.e. it reveals a
linear dependence on $u$ with a slope $\gamma(r)$ given by eq. 
(\ref{GammaEinDimInk}). With respect to the works \cite{PRLTurb, PhysicaD, 
Yakhot, Davoudi} it remains open to explain why $\gamma$ deviates from 
the value $1/3$.

A further new important result is that the drift coefficient
$D^{(1)}_{x}$ does not depend on the velocity increment $u$ and is a
linear function of the logarithmic energy dissipation rate $x$ as
specified by eq.  (\ref{D1xVonXFunk}).  One thus obtains a system of
two stochastic differential equations for the evolution of $u$ and $x$
in $r$ which are linked only via their stochastic terms while the
deterministic parts of the equations for $u$ and $x$ do not depend on 
the other variable.

To summarize, the Langevin--equation for the two dimensional stochastic
variable ${\mathbf q}(r) = \left( \, u(r), x(r) \, \right)$ reads:
\begin{eqnarray}
    - \frac{\partial}{\partial r} \, u(r) \; & = & \; - \frac{1}{r} \,
    \gamma(r) \, u(r) \; + \; m \, \exp\left( \frac{a_{1}}{2} x(r)
    \right) \; \Gamma_{u}(r) , \nonumber \\
    - \frac{\partial}{\partial r} \, x(r) \; & = & \; + \frac{1}{r} \,
    G(r) \, x(r) \; + \; \frac{1}{r} \, F(r) \; + \; \sqrt{\,
    \frac{1}{r} \, D^{(2)}_{xx}(u,x,r) \, } \; \Gamma_{x}(r) \, . 
    \label{TwoDimLangevinResult}
\end{eqnarray}
According to eq.  (\ref{D2uuValues}), the coefficient $m =
\sqrt{\frac{a_{0}(r)}{r}}$ in the stochastic part of the equation for
$u(r)$ is approximately constant in $r$ (see also fig. 
\ref{D2uuKoeffs}).  Note also that, when rewritten in terms of
$\epsilon_{r}$, the stochastic term in the equation for $u(r)$
exhibits a simple power--law dependence on the energy dissipation
rate: $\exp \left( a_{1}x/2\right) = \left(
\epsilon_{r}/\bar{\epsilon} \right)^{a_{1}/2} \approx
\sqrt{\epsilon_{r}/\bar{\epsilon}}$ (see fig.  \ref{D2uuKoeffs}b).

It is interesting to compare the two dimensional equation
(\ref{TwoDimLangevinResult}) with the one dimensional
Langevin--equations for $u(r)$ and $x(r)$ discussed in section
\ref{eindim}. Those equations read:
\begin{eqnarray}
    - \frac{\partial}{\partial r} \, u(r) \; & = & \; - \frac{1}{r} \,
    \gamma(r) \, u(r) \; + \; \sqrt{ \, \frac{1}{r} \, \left( \,
    \alpha(r) - \delta(r) u + \beta(r) u^2 \, \right) \, } \;
    \Gamma(r) \, , \nonumber \\
    - \frac{\partial}{\partial r} \, x(r) \; & = & \; + \frac{1}{r} \,
    G(r) \, x(r) \; + \; \frac{1}{r} \, F(r) \; + \; \sqrt{\,
    \frac{1}{r} \, D(r) \, } \; \Gamma(r) \, . 
    \label{OneDimLangevinResult}
\end{eqnarray}
Compared to the one dimensional Langevin--equation, the stochastic
part of the equation for $u(r)$ is considerably simpler for the two
dimensional process.  Note also that the resulting equation for the
evolution of $u(r)$ in two dimensions is symmetric in $u \rightarrow
-u$: if $x$ was constant, the solutions of eq. 
(\ref{TwoDimLangevinResult}) would be symmetric in $u$, thus leading
to vanishing odd order moments.  Asymmetries of the velocity
increment, which are of importance in Kolmogorov's four--fifths law,
can thus clearly be attributed to the influence of the energy
cascade. 

While the equation for the velocity increment simplifies considerably
in the two dimensional formulation, the diffusion coefficient
$D^{(2)}_{xx}(u,x,r)$ seems to exhibit a more complex dependence on
its arguments (see fig.  \ref{D2xx}) than the one dimensional
coefficient $D(r)$.  Whether the dependence of $D^{(2)}_{xx}$ on $x$
is in fact given by the rather complex form indicated in fig. 
\ref{D2xx}(b) is yet an open problem which requires further
discussions of the experiemental uncertainties.  However, it is also
conceivable that an even more complete characterization of the
turbulent cascade has to include the transversal velocity increment as
well.  Such a three--dimensional analysis might lead to further
simplifications of the diffusion terms.

Nevertheless, the results obtained so far give reason to believe that
the mathematical framework of stochastic Markovian processes is a
suitable tool for experimental investigations concerning the joint
statistical properties of velocity increments and the energy
dissipation rate in fully developed turbulence.  As it is known that
the Fokker--Planck equation holds for the conditional pdf as well, the
knowledge of this equation also provides all information about any
N--scale joint pdf of $x$ and $u$, see equations \ref{chain} and
\ref{FoplaCond}.  Thus any general moment of
$u(r)^{\alpha}x(r')^{\beta}$ is known.  In this way the
phenomenological Fokker--Planck equation derived from the data
provides a closed description for all moments on all scales.

\section*{Acknowledgements} 

We gratefully acknowledge fruitful discussions with M. Siefert, St. 
L\"uck,  A. Naert, and 
P. Marq. We also acknowledge discussions with B. Chabaud and O. Chanal
from whom we got the data.

\end{document}